\documentclass[twocolumn,prl,superscriptaddress]{revtex4}
\usepackage{amsmath,amssymb,amsfonts,mathrsfs}
\usepackage{bm}
\usepackage{graphicx}
\usepackage{hyperref}
\usepackage{amsbsy}
\usepackage[english]{babel}
\usepackage{dcolumn}

\hypersetup{
     colorlinks = true,
     citecolor  = red,  
     linkcolor  = blue  
}

\begin{document}

\title{Experimental Measurement of the Quantum Metric Tensor and Related Topological Phase Transition with a Superconducting Qubit}

\author{Xinsheng Tan}
\email{meisen0103@163.com}
\affiliation{National Laboratory of Solid State Microstructures, School of Physics,
Nanjing University, Nanjing 210093, China}
\author{Dan-Wei Zhang}
\email{danweizhang@m.scnu.edu.cn}
\affiliation{Guangdong Provincial Key Laboratory of Quantum Engineering and Quantum Materials, GPETR Center for Quantum Precision
Measurement and SPTE, South China Normal University, Guangzhou 510006, China}

\author{Zhen Yang}
\affiliation{National Laboratory of Solid State Microstructures, School of Physics,
Nanjing University, Nanjing 210093, China}
\author{Ji Chu}
\affiliation{National Laboratory of Solid State Microstructures, School of Physics,
Nanjing University, Nanjing 210093, China}
\author{Yan-Qing Zhu}
\affiliation{National Laboratory of Solid State Microstructures, School of Physics,
Nanjing University, Nanjing 210093, China}
\author{Danyu Li}
\affiliation{National Laboratory of Solid State Microstructures, School of Physics,
Nanjing University, Nanjing 210093, China}
\author{Xiaopei Yang}
\affiliation{National Laboratory of Solid State Microstructures, School of Physics,
Nanjing University, Nanjing 210093, China}
\author{Shuqing Song}
\affiliation{National Laboratory of Solid State Microstructures, School of Physics,
Nanjing University, Nanjing 210093, China}
\author{Zhikun Han}
\affiliation{National Laboratory of Solid State Microstructures, School of Physics,
Nanjing University, Nanjing 210093, China}
\author{Zhiyuan Li}
\affiliation{National Laboratory of Solid State Microstructures, School of Physics,
Nanjing University, Nanjing 210093, China}
\author{Yuqian Dong}
\affiliation{National Laboratory of Solid State Microstructures, School of Physics,
Nanjing University, Nanjing 210093, China}

\author{Hai-Feng Yu}
\affiliation{National Laboratory of Solid State Microstructures, School of Physics,
Nanjing University, Nanjing 210093, China}
\author{Hui Yan}
\affiliation{Guangdong Provincial Key Laboratory of Quantum Engineering and Quantum
Materials, School of Physics and Telecommunication Engineering, South China Normal University, Guangzhou 510006, China}
\author{Shi-Liang Zhu}
\email{slzhu@nju.edu.cn}
\affiliation{National Laboratory of Solid State Microstructures, School of Physics,
Nanjing University, Nanjing 210093, China}
\affiliation{Guangdong Provincial Key Laboratory of Quantum Engineering and Quantum Materials, GPETR Center for Quantum Precision
Measurement and SPTE, South China Normal University, Guangzhou 510006, China}
\author{Yang Yu}
\email{yuyang@nju.edu.cn}
\affiliation{National Laboratory of Solid State Microstructures, School of Physics,
Nanjing University, Nanjing 210093, China}

\begin{abstract}
Berry curvature is an imaginary component of the quantum geometric tensor (QGT) and is well studied in many branches of
modern physics; however, the quantum metric as a real component of the QGT is less explored. Here, by using tunable superconducting circuits, we experimentally demonstrate two methods to directly measure the quantum metric tensor for characterizing the geometry and topology of underlying quantum states in parameter space. The first method is to probe the transition probability after a sudden quench, 
and the second one is to detect the excitation rate under weak periodic driving. Furthermore, based on quantum-metric and Berry-curvature measurements, we explore a topological phase transition in a simulated time-reversal-symmetric system, which is characterized by the Euler characteristic number instead of the Chern number. The work opens up a unique approach to explore the topology of quantum states with the QGT.
\end{abstract}

\maketitle

\bigskip

\emph{Introduction.}-- Geometry is one of the most intriguing concepts for understanding diverse fundamental phenomena in modern physics, ranging from general relativity and gauge theory to quantum mechanics. The geometry of quantum states in Hilbert space can be described by the so-called quantum geometric tensor (QGT) \cite{Kolodrubetz2017,Provost1980,Xiao2010}, whose real and imaginary components define the quantum metric (or Fubini-Study metric) and the Berry curvature, respectively. The Berry curvature has been explored in various topics including the celebrated Aharonov-Bohm effect \cite{AB}, Berry phase effect \cite{Xiao2010,Berry}, and the more recent topological quantum matter \cite{Hasan,Qi2011,Armitage,DWZhang2018}. For instance, the integral of the Berry curvature over a closed two-dimensional manifold defines the first Chern number \cite{TKNN}, which is a topological invariant to characterize quantized Hall conductivity and Chern insulators \cite{Haldane}. Recently, the Berry curvature has been directly measured in some engineered systems, such as cold atoms \cite{ColdAtom1,ColdAtom2,ColdAtom3,ColdAtom4,DWZhang2018}, photonic lattices \cite{Wimmer2017}, and superconducting quantum circuits \cite{Schroer2014,Roushan2014,Tan2018,Tan2019}.

Compared to the well-studied Berry curvature, the quantum metric is less explored, both theoretically and experimentally.
The quantum metric is an intrinsic geometric concept that reflects the distance between neighbouring quantum states in parameter space and is associated with a different topological invariant, the Euler characteristic number (ECN), for describing the topology of the state manifold \cite{Anandan1990,Ma2010,Rezakhani,Ma2013,Kolodrubetz2013}. Recently, it was shown that the quantum metric is closely related to the quantum phase transition \cite{Zanardi2007}, superfluidity in flat bands \cite{Julku2016}, and topological matter \cite{Roy2014,Lim2015,Palumbo2018}. Several schemes have been proposed to extract the quantum metric \cite{Kolodrubetz2013,Neupert2013,Ozawa2018a,Ozawa2018b,Bleu2018a,Bleu2018b,Klees2018}, but its direct experimental measurement remains elusive. Moreover,  to the best of our knowledge, a topological phase transition characterized by the ECN \cite{Ma2013}, which can be obtained from the quantum-metric measurements, is still lacking in experiments.

In this Letter, we experimentally demonstrate two different methods to directly measure the QGT in parameter space using tunable superconducting circuits. The first method is to probe the transition probability after a sudden quench of the system parameters, which is associated with the wave-function overlap. The second method is to detect the excitation rate under weak periodic driving on a quantum state, as proposed in Ref. \cite{Ozawa2018a}. Furthermore, based on quantum-metric and Berry-curvature measurements, we explore a phase transition in a simulated time-reversal-symmetric (TRS) system, which is characterized by the ECN instead of the (vanishing) Chern number. Our work thus provides generic and versatile tools to reveal the geometry and topology of quantum states.

\begin{figure}[tbp]
\centering
\includegraphics[width=7cm]{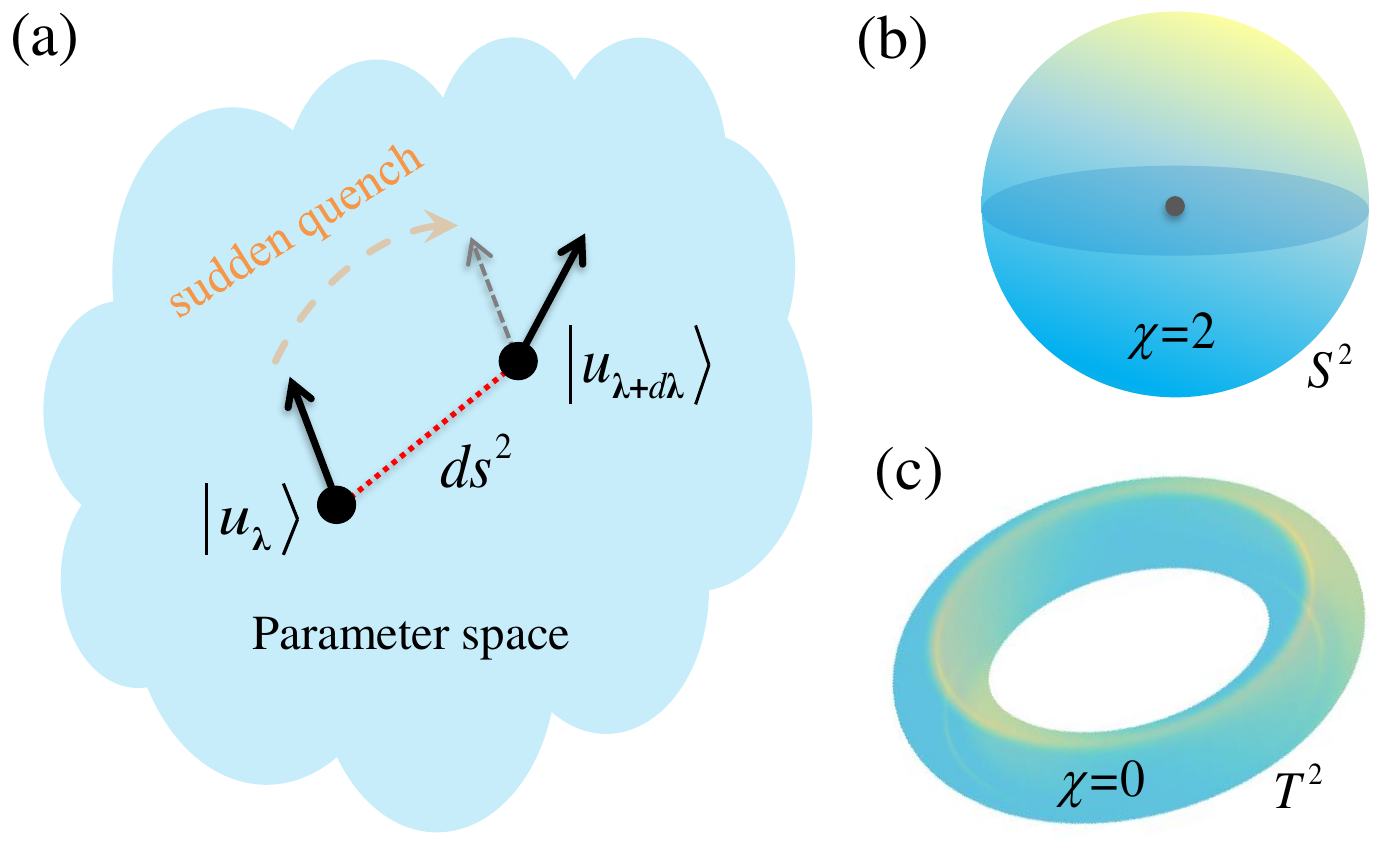}
\caption{(Color online) (a) Illustration of distance between two nearby quantum states in parameter space. The topology of a closed manifold can be characterized by the ECN $\chi$ in Eq. (\ref{EulerNum}): (b) $\chi=2$ for a sphere $S^2$; and (c) $\chi=0$ for a torus $T^2$.}
\label{demon}
\end{figure}

\emph{Quantum metric and measurement schemes.}-- We consider a generic (non-degenerate) quantum state $|u_{\boldsymbol\lambda}\rangle$ in a parameter space spanned by $N$ dimensionless parameters $\boldsymbol\lambda = (\lambda_1, \lambda_2, \cdots, \lambda_N)$. The so-called QGT associated with $|u_{\boldsymbol\lambda}\rangle$ is defined as \cite{Kolodrubetz2017}
\begin{equation}
Q_{\mu\nu}=\langle \partial_{\lambda_{\mu}} u_{\boldsymbol\lambda} |(1 - | u_{\boldsymbol\lambda}\rangle \langle u_{\boldsymbol\lambda}|)| \partial_{\lambda_{\nu}}u_{\boldsymbol\lambda} \rangle=g_{\mu\nu}+i\mathcal{F}_{\mu\nu}/2,
\end{equation}
which can take complex values. Its real component defines the quantum metric $g_{\mu\nu} = \mathrm{Re}[Q_{\mu\nu}]=g_{\nu\mu}$, which is symmetric. The imaginary component is related to the Berry curvature $\mathcal{F}_{\mu\nu} =2\mathrm{Im}[Q_{\mu\nu}]=-\mathcal{F}_{\nu\mu}$, which is antisymmetric. The quantum metric defines a distance between nearby states $|u_{\boldsymbol\lambda}\rangle$ and $|u_{\boldsymbol\lambda+d\boldsymbol\lambda}\rangle$ in parameter space \cite{Provost1980,Supp}
\begin{equation}\label{ds}
ds^2=1-|\langle u_{\boldsymbol\lambda}| u_{\boldsymbol\lambda+d\boldsymbol\lambda}\rangle|^2=\sum\nolimits_{\mu\nu} g_{\mu\nu} d\lambda_{\mu} d\lambda_{\nu},
\end{equation}
which is related to the wave-function overlap, as shown in Fig. \ref{demon}(a). The quantum metric and the Berry curvature are gauge-invariant and describe the complete geometry of the quantum state manifold.

We take a two-level system in a Bloch sphere $S^2$ (two-dimensional Hilbert space) as an example to illustrate the basic ideas of the QGT. The system Hamiltonian can be parameterized as
\begin{align}
	H_1(\theta, \phi)
	=
	\frac{\Omega}{2}
	\begin{pmatrix}
	\cos \theta & \sin \theta e^{-i\phi} \\
	\sin \theta e^{i\phi} & - \cos \theta
	\end{pmatrix} , \label{Ham1}
\end{align}
where the two angles $\{\theta,\phi\}$ represent the parameters $\boldsymbol\lambda$. For both the ground state $|u_-\rangle=(\sin\frac{\theta}{2},-\cos\frac{\theta}{2}e^{i\phi})^{\text{Tr}}$ and the excited state $|u_+\rangle=(\cos\frac{\theta}{2},\sin\frac{\theta}{2}e^{i\phi})^{\text{Tr}}$ with $\text{Tr}$ denoting the transpose of a matrix, the quantum metric is given by $g_{\theta \theta} = 1/4$, $g_{\phi \phi} = \sin^2 \theta/4$, and $g_{\theta \phi}=0$. However, the Berry curvature for $|u_{\pm}\rangle$ is the opposite: $\mathcal{F}_{\theta\phi}^{\pm}=\pm\frac{1}{2}\sin\theta$. By viewing the Berry connection as the vector potential, one simulates a Dirac monopole at the center of the Bloch sphere [cf. Fig. \ref{demon}(b)], which has a topological charge in terms of the first Chern number $C_{\pm}=\frac{1}{2\pi}\int_{S^2}\mathcal{F}_{\theta\phi}^{\pm}d\theta d\phi=\pm1$. The topology of the two-dimensional manifold $\mathcal{M}$ can also be characterized by another topological invariant related to the metric tensor, i.e., the ECN \cite{Ma2013,Kolodrubetz2013}
\begin{equation}
\chi=\frac{1}{4\pi}\int_{\mathcal{M}} R \sqrt{\det g}~d\mu d\nu, \label{EulerNum}
\end{equation}
where $R$ is the Ricci scalar curvature and $g$ denotes the $2\times2$ metric tensor. For $\mathcal{M}=S^2$ with $\{\mu,\nu\}=\{\theta,\phi\}$ and $R=8$, $\chi=2$ for both $|u_{\pm}\rangle$. Here $\chi=2|C_{\pm}|$ due to an intrinsic relation between the Berry curvature associated with the monopole and the determinant of the metric tensor: $F_{\theta\phi}=2\sqrt{\det g}$. If the closed manifold is a torus $\mathcal{M}=T^2$ without monopoles, one has $\chi=0$, as shown in Fig. \ref{demon}(c).


Different from the first Chern number which should vanish in a system in the presence of TRS, the ECN $\chi$ can be nontrivial under this condition. Below we will show that this index can characterize a topological phase transition in a simulated TRS system: $\chi$ rapidly jumps with respect to an external parameter, while the Chern number is always zero and thus not an appropriate topological index.

The distance $ds^2$ in Eq. (\ref{ds}) actually gives the transition probability $P^+=ds^2$ of the quantum state being excited to other eigenstates during a sudden quench of the parameter from $\boldsymbol\lambda$ to $\boldsymbol\lambda+\delta\boldsymbol\lambda$ \cite{Kolodrubetz2017,Lim2015}, as shown in Fig. \ref{demon}(a). This observation provides a generic and direct approach to measure the quantum metric tensor via the transition probability in the sudden approximation. We initially prepare the system  at $\boldsymbol\lambda=\boldsymbol\lambda_0$  to detect the quantum metric at that point. To extract the diagonal components $g_{\mu\mu}$, we suddenly quench the system parameter to $\boldsymbol\lambda_0+\delta\lambda \boldsymbol e_{\mu}$ along the $\boldsymbol e_{\mu}$ direction and then measure the transition probability $P^+_{\mu\mu}=g_{\mu\mu}\delta\lambda^2+\mathcal{O}(\delta\lambda^3)$. To extract the off-diagonal components $g_{\mu\nu}$ ($\mu\neq\nu$), we apply a sudden quench to $\boldsymbol\lambda_0+\delta\lambda \boldsymbol e_{\mu}+\delta\lambda\boldsymbol e_{\nu}$ along the $\boldsymbol e_{\mu}+\boldsymbol e_{\nu}$ direction and then measure the probability $P^+_{\mu\nu}$, which has the relation $P^+_{\mu\nu}-P^+_{\mu\mu}-P^+_{\nu\nu}=2g_{\mu\nu}\delta\lambda^2+\mathcal{O}(\delta\lambda^3)$. Another scheme we used to extract the components of the quantum metric is via the excitation rate of a quantum state upon applying proper time-periodic modulations \cite{Ozawa2018a,Supp}.

\emph{Experimental system.}-- We realize a highly tunable two-level Hamiltonian with superconducting quantum circuits and measure the quantum metric using the two different methods. The circuits consist of a superconducting transmon qubit embedded in a three-dimensional aluminium (Al) cavity~ \cite{Tan2018,Tan2019,paik_3d,devoret_3d,JinXY,DiCarlo}. The transmon qubit is composed of an Al single Josephson junction and two pads (250 $\mu $m $\times $ 500 $\mu$m) fabricated on a 500 $\mu $m thick silicon substrate. The thicknesses of the two Al films of the junction are 30nm and 80nm, respectively. The cavity in our experiments is mainly used to  conveniently control and readout the transmon qubit. The resonance frequency of the cavity TE101 mode is 9.053 GHz. The whole sample package is cooled in a dilution refrigerator to a base temperature of 10 mK. The experimental setups for the qubit control and measurement are well established \cite{Tan2018,Tan2019,paik_3d,devoret_3d,JinXY,DiCarlo}.
The coupled transmon qubit and cavity exhibit anharmonic multiple energy levels. In our experiments, we use the lowest two energy levels $ |0\rangle $ and $|1\rangle$, which are coupled with the microwave fields. The transition frequency between them is  $ \omega _{10}/2\pi = $ 7.172 GHz, which is independently determined by saturation spectroscopies. The energy relaxation time of the qubit is $T_1\sim$7 $\mu s$, and the dephasing time is $T^*_2\sim$6 $\mu s$. We apply the microwave driving along $x$, $y$, and $z$ directions and realize the effective Hamiltonian in the rotating frame (let $\hbar =1$)
\begin{equation}
H_{\text{exp}}=\frac{1}{2}(\Omega_x\sigma_x+\Omega_y\sigma_y+\Omega_z\sigma_z), \label{Ham_qubit}
\end{equation}
where $\sigma_{x,y,z}$ are the three Pauli matrices. Here $\Omega _{x}$ $(\Omega _{y})$ is the Rabi frequency along the $x$ ($y$) axis on the Bloch sphere, which can be controlled by tuning the amplitude and phase of the microwave applied to the system, $\Omega _{z}=\omega _{10}-\omega_m$ is determined by the detuning between the system energy level spacing $\omega _{10}$ and the microwave frequency $\omega_m$. The parameters $\Omega_{x,y,z}$ are calibrated by using Rabi oscillations and Ramsey fringes in our experiments \cite{Supp}.

\begin{figure}[tbph]
\includegraphics[width=8cm]{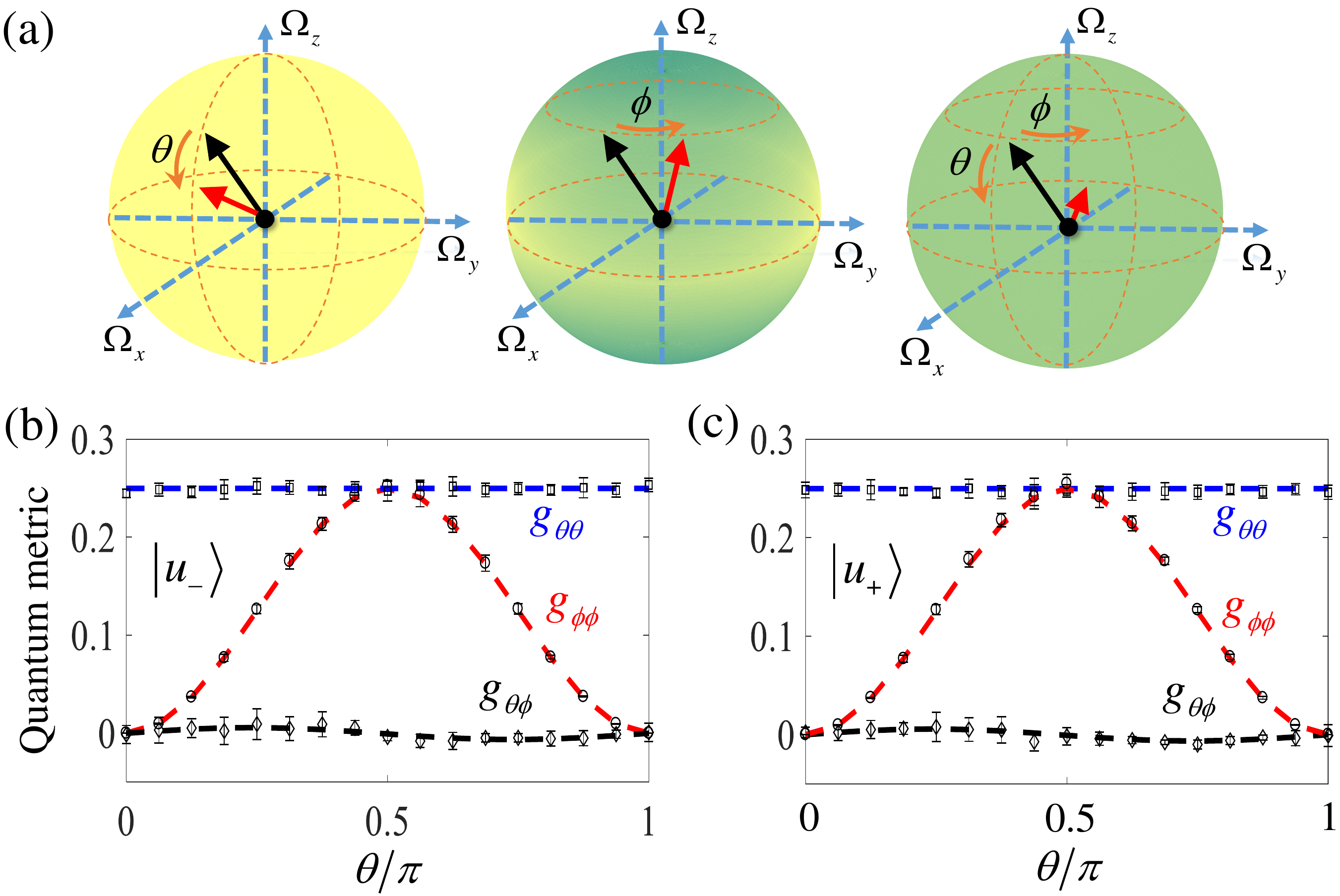}
\caption{(Color online) Measuring the quantum metric from transition probability after sudden quench. (a) Three quench protocols for extracting $g_{\theta\theta}$, $g_{\phi\phi}$, and $g_{\theta\phi}$ (from left to right).
Measured $g_{\mu\nu}$ as a function of $\theta$ for (b) the ground state $|u_-\rangle$ and (c) the excited state $|u_+\rangle$. Dashed lines are numerical results from full-time-dynamics simulations.}
\label{QuenchExp}
\end{figure}

In our experiments, we can simultaneously tune the frequency, amplitude, and phase of the microwaves to create arbitrary two-level Hamiltonians. In particular, we work with two Hamiltonians that can be represented as a sphere and a torus manifold in parameter space, respectively. First of all, we realize $H_{\text{exp}}=H_1(\theta,\phi)$ in Eq. (\ref{Ham1}) by setting the microwave point by point to  design the three parameters as $\Omega_x=\Omega \sin\theta\cos\phi$, $\Omega_y=\Omega \sin\theta\sin\phi$, and $\Omega_z=\Omega \cos\theta$, where $\Omega=10$ MHz is set as the energy unit.

\emph{Measuring the quantum metric by sudden quench.}-- We now describe our experiments to measure the quantum metric $g_{\theta\theta}$, $g_{\phi\phi}$ and $g_{\theta\phi}$ by using the sudden quench scheme. The three quench protocols are illustrated in Fig. \ref{QuenchExp}(a). The system is initially prepared at an eigenstate of the Hamiltonian $H_1(\boldsymbol\lambda_0)$ in the parameter space $\boldsymbol\lambda=\{\theta,\phi\}$. Then the Hamiltonian is rapidly swept to $H_1(\boldsymbol\lambda_0+\delta\boldsymbol\lambda)$, followed by {\color{red}a} state tomography to obtain the transition probability. In our experiment, we set the parameter quench $\boldsymbol\lambda(t)=\boldsymbol\lambda_0+t/T\delta\lambda\boldsymbol e_{\eta}$ along the $\boldsymbol e_{\eta}$ direction, where the quench time $T=5$ ns \cite{note1}, $\delta \lambda=\pi/16$, and $\boldsymbol e_{\eta}=\{\boldsymbol e_{\theta}, \boldsymbol e_{\phi}, \boldsymbol e_{\theta}+\boldsymbol e_{\phi}\}$ for the three protocols, respectively.
\begin{figure}[tbph]
\includegraphics[width=8cm]{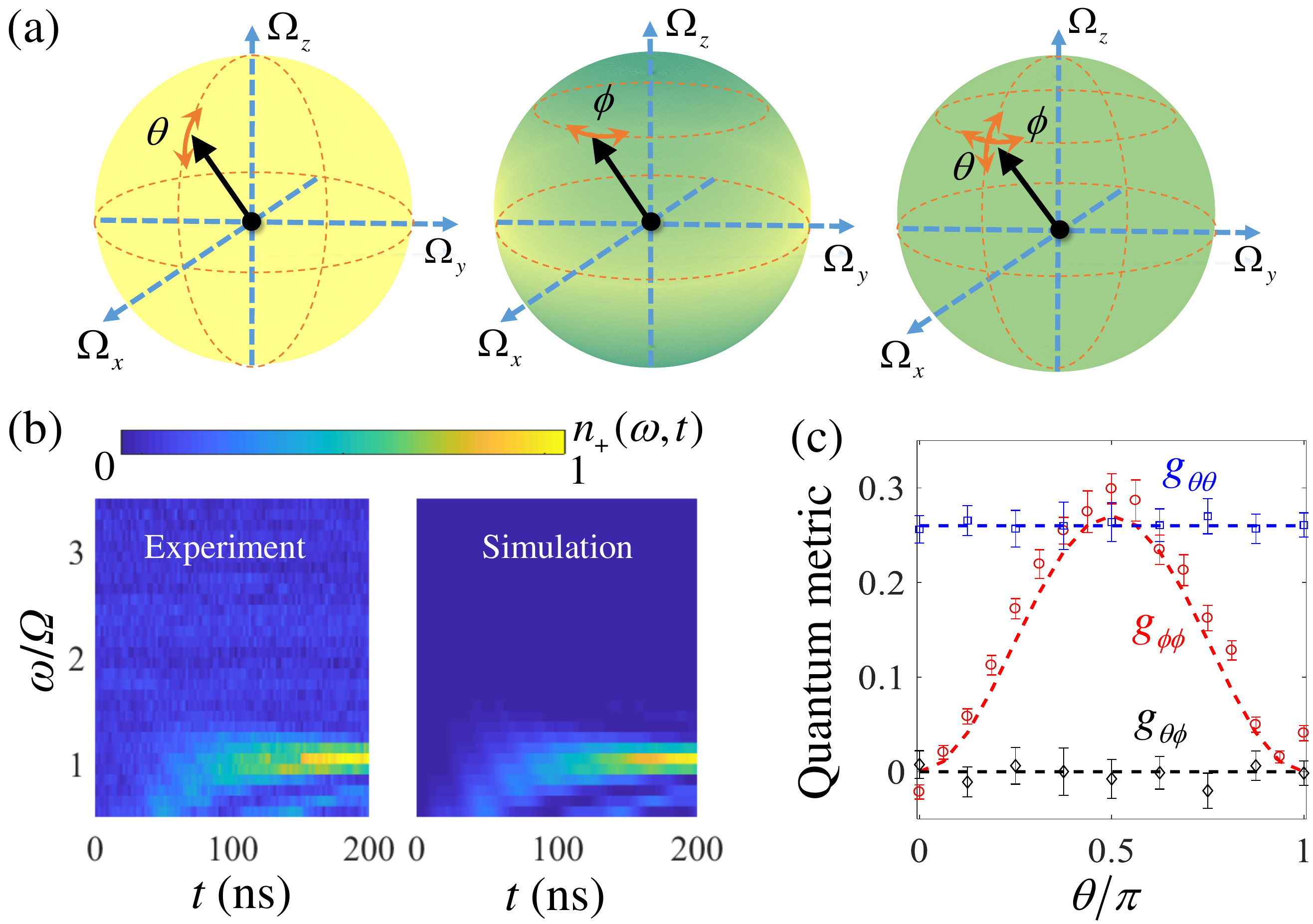}
\caption{(Color online) Measuring the quantum metric from excitation rate under periodic weak drive for the ground state $|u_-\rangle$. (a) Three drive protocols for extracting $g_{\theta\theta}$, $g_{\phi\phi}$, and $g_{\theta\phi}$ (from left to right). 
(b) Measured and simulated results of excitation probability $n_+$ on $|u_+\rangle$ as a function of the $\theta$-driving frequency $\omega$ and time $t$, with the initial state at $\theta_0=\pi/2$ and $\phi_0=0$. (c) Extracted $g_{\mu\nu}$ as a function of $\theta$ for the measurement time $t_{\text{meas}}=200$ ns. Dashed lines in (c) are numerical results from full-time-dynamics simulations. The modulation amplitude $E/\omega=0.1$ is fixed in the experiment.}
\label{DriveExp}
\end{figure}
This means that only $\theta$ or $\phi$ is ramping linearly in the first two quenches, while $\theta$ and $\phi$ are simultaneously ramping in the third quench. From state tomography, we can extract the quantum metric at $\boldsymbol\lambda_0$ from the measured transition probability: $g_{\theta\theta}\approx P^+_{\theta\theta}/\delta \lambda^2$, $g_{\phi\phi}\approx P^+_{\phi\phi}/\delta \lambda^2$, and $g_{\theta\phi}\approx(P^+_{\theta\phi}-P^+_{\theta\theta}-P^+_{\phi\phi})/2\delta \lambda^2$. In order to simplify the state initialization and increase the measurement fidelity, we rotate the frame axis in the experiment to keep the eigenstate of the initial Hamiltonian always in the energy level $|0\rangle$ \cite{Supp}. The measured quantum metric $g_{\mu\nu}$ as a function of $\theta$ for the ground state $|u_-\rangle$ and the excited state $|u_+\rangle$ are respectively shown in Fig. \ref{QuenchExp}(b) and \ref{QuenchExp}(c), which agree well with the numerical results from full-time-dynamics simulations. The obtained ECN from Eq. (\ref{EulerNum}) for $|u_-\rangle$ ($|u_+\rangle$) is $\chi=1.99\pm0.04$ ($\chi=2.01\pm0.04$), which is close to the theoretical value $2$. In order to obtain the full QGT, we also measure the Berry curvature $\mathcal{F}_{\theta\phi}$ using the quasi-adiabatic dynamical response method \cite{Schroer2014,Roushan2014,Tan2018,Tan2019,Gritsev2012,Supp}, and extract the Chern number $C=1.05\pm0.10$ ($C=0.96\pm0.06$) for $|u_-\rangle$ ($|u_+\rangle$).

\emph{Measuring the quantum metric by periodic drive.}-- We then demonstrate another method to measure the quantum metric of the ground state $|u_-\rangle$ from excitation rate under periodic weak drive \cite{Ozawa2018a,Supp}. The implemented three drive protocols for extracting $g_{\theta\theta}$, $g_{\phi\phi}$, and $g_{\theta\phi}$ are illustrated in Fig. \ref{DriveExp}(a). The system is initially prepared at $\boldsymbol\lambda_0=(\theta_0,\phi_0)$, and only the parameter $\theta$ ($\phi$) is modulated as $\theta(t)=\theta_0+2(E/\omega) \cos(\omega t)$ (with $\theta\leftrightarrow\phi$) to detect $g_{\theta\theta}$ ($g_{\phi\phi}$) at that point, with frequency $\omega$ and small modulation amplitude $E/\omega\ll1$. Under this modulation, the system will be excited to $|u_+\rangle$ with the probability $n_+$. An example of measured and simulated $n_+(\omega,t)$ as a function of $\omega$ and the evolution time $t$ is shown in Fig. \ref{DriveExp}(b). In our experiment, we set the modulation amplitude $E/\omega=0.1$ and the frequency range $\omega\in[\omega_{\text{min}},\omega_{\text{max}}]$, where $\omega_{\text{min}}=0.5\Omega$ and $\omega_{\text{max}}=3.5\Omega$. At the measurement time $t_{\text{meas}}=200$ ns, we can obtain the excitation rate $\Gamma(\omega)=n_+(\omega,t_{\text{meas}})/t_{\text{meas}}$. Based on the time-dependent perturbation theory \cite{Ozawa2018a}, the diagonal component of the quantum metric $g_{\theta\theta}$ ($g_{\phi\phi}$) can be obtained from the equation: $g_{\theta\theta}=\int_{\omega_{\text{min}}}^{\omega_{\text{max}}} \Gamma(\omega) d\omega/(2\pi E^2)$ ($\theta\leftrightarrow\phi$). Similarly, the off-diagonal component $g_{\theta\phi}$ can be extracted from the excitation rate \cite{Ozawa2018a,Supp}, by combining two sets of measurements while simultaneously modulating the two parameters $\theta(t)=\theta_0+2(E/\omega) \cos(\omega t)$ and $\phi(t)=\phi_0\pm2(E/\omega) \cos(\omega t)$. Finally, the measured quantum metric as a function of $\theta$ is shown in Fig. \ref{DriveExp}(c), which agrees with the full-time-dynamics simulation results. The obtained ECN is $\chi=2.38\pm0.30$. The uncertainty here is much larger than that of the sudden quench. The reason is that for the periodic drive, before evolving the qubit and performing tomography, we cannot evade the state initialization by rotating the frame axis as in the sudden quench. The imperfection of state preparation due to the qubit control and decoherence significantly increase the fluctuation.

\emph{Topological phase transition characterized by the ECN.}-- To further show the application of the quantum metric, we explore phase transition characterized by the ECN in our superconducting circuits. By designing the microwave fields on the qubit, we realize the second two-level Hamiltonian proposed in Ref. \cite{Ma2013} on a manifold spanned by the parameters $\boldsymbol k=(k_x,k_y)$ with $\{k_x,k_y\}\in[0,2\pi]$:
\begin{equation}
H_{\text{exp}}=H_2(\boldsymbol k)=\frac{\Omega}{2}
\begin{pmatrix}
	-h-\cos k_x   &   \alpha\sin k_x e^{-ik_y}\\
   \alpha\sin k_x e^{ik_y}   & h+\cos k_x
\end{pmatrix}, \label{Ham2}
\end{equation}
where $h$ is a tunable, dimensionless parameter and $\alpha=0.5$ in our experiment.
The Hamiltonian $H_2(\boldsymbol k)$ can describe a Bloch band in momentum ($\boldsymbol k$) space for spinless free fermions. It is equivalent to the many-body Hamiltonian of the XY spin chain after the Jordan-Wigner and Fourier transformations to a (1+1)-dimensional momentum space \cite{Ma2013,Carollo2005,SLZhu2006,Sachdev1999}, which is spanned by the reduced $k_x\in[0,\pi]$ and a local $U(1)$ gauge phase $\varphi\in[0,2\pi]$ acting as $k_y$. Under the time reversal operator $\mathcal{T}=\mathcal{K}$, where $\mathcal{K}$ is the complex conjugation, one can find that $H_2(\boldsymbol k)$ is TRS since it satisfies the condition $\mathcal{T}H_2(\boldsymbol k)\mathcal{T}^{-1}=H_2(\boldsymbol k)$.
Due to the TRS, the Berry curvature $\mathcal{F}(\boldsymbol k)=-\mathcal{F}(-\boldsymbol k)$ \cite{Supp}, which gives the zero Chern number as the integral of $\mathcal{F}(\boldsymbol k)$ in the torus $T^2$. Thus, we need an appropriate topological number to characterize the phase transition for this case. We demonstrate that the ECN $\chi$ acts as such an index. In order to obtain $\chi$ in Eq. (\ref{EulerNum}), we calculate $\sqrt{\det g}$ for the ground state $|u_{\boldsymbol k}\rangle$ \cite{Supp}: $$\sqrt{\det g}=\alpha^2|(1+h\cos k_x)\sin k_x|/(2f^{3/2})$$ with $f=(h+\cos k_x)^2+\alpha^2\sin^2 k_x$, which is independent of $k_y$. Thus, one can obtain $\chi=4\int_{0}^{2\pi}\sqrt{\det g} dk_x$ under the condition $\{k_x,k_y\}\in[0,2\pi]$, while the integration over $k_x$ is taken from $0$ to $\pi$ for the spin chain model discussed in Ref. \cite{Ma2013}. This gives rise to the difference of $\chi$ with a factor $2$ between the two cases, which can be intuitionally understood as the doubling of the monopoles in the enlarged state manifold.

\begin{figure}[tbph]
\includegraphics[width=7.5cm]{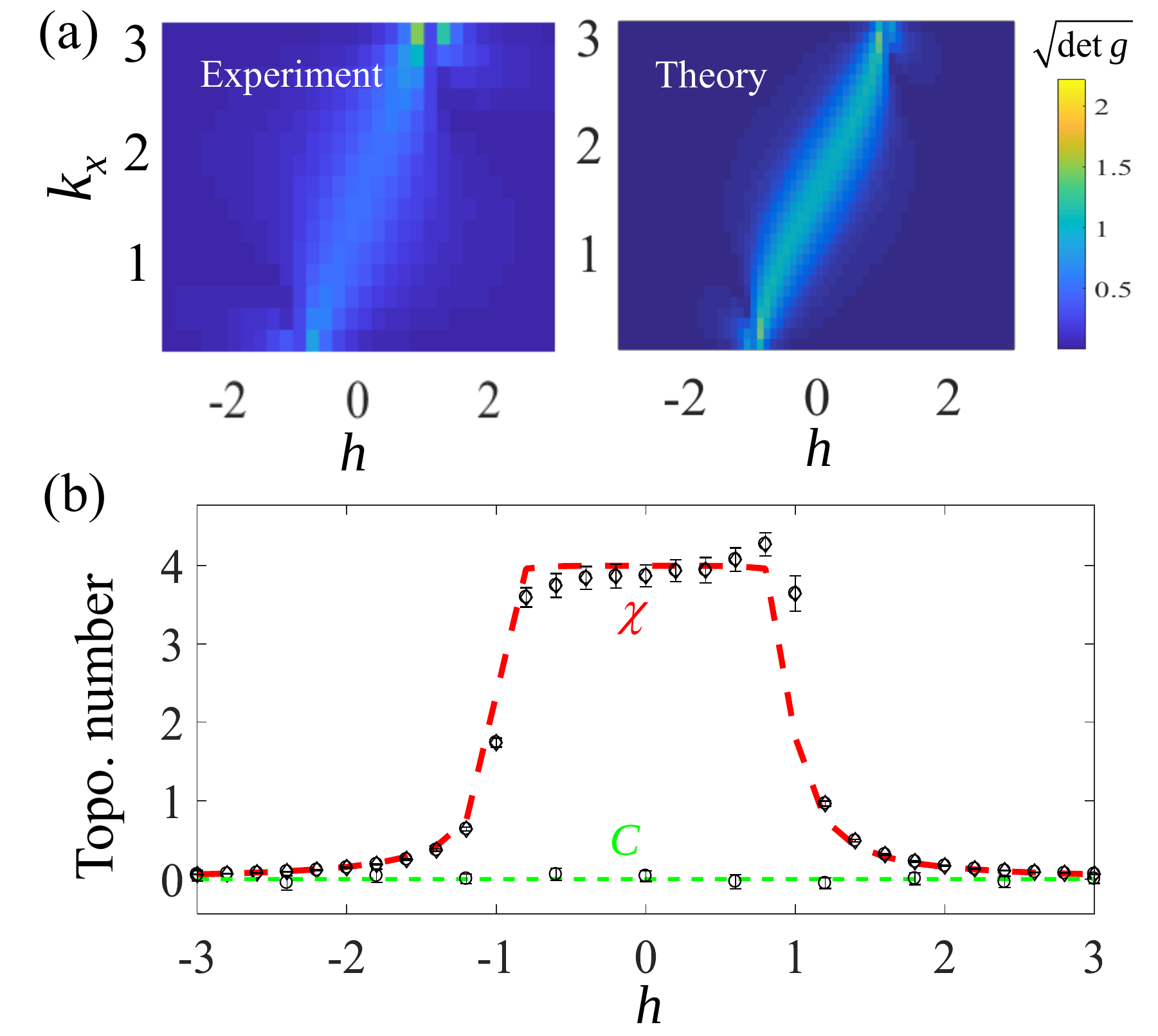}
\caption{(Color online) Characterizing phase transition by the ECN $\chi$ for the simulated Hamiltonian in Eq. (\ref{Ham2}). (a) Measured (left) and theoretical (right) results of $\sqrt{\det g}$ as a function of parameters $h$ and $k_x$. (b) The obtained $\chi$ and $C$ (the Chern number) as functions of $h$. Here $\chi\approx4$ declines rapidly to $\chi\approx0$ near $h=\pm1$, showing the phase transition, while $C\approx0$ for all $h$ due to the TRS.}
\label{TopTrans}
\end{figure}	

In our experiment, we first prepare the qubit in the ground state of the simulated Hamiltonian $H_2(\boldsymbol k)$ by setting the microwave with the three parameters as $\Omega_x=\alpha\Omega \sin k_x\cos k_y$, $\Omega_y=\alpha\Omega \sin k_x\sin k_y$, and $\Omega_z=\Omega (h+\cos k_x)$, where $\Omega=10$ MHz is set as the energy unit. Then, using the sudden-quench protocols, we directly measure the quantum metric. The measured value of $\sqrt{\det g}$ as a function of the parameters $k_x$ and $h$ is plotted in Fig. \ref{TopTrans}(a) (left), which agrees with the theoretical result (right). The extracted $\chi$ as a function of $h$ is shown in Fig. \ref{TopTrans}(b). It is clear that $\chi\approx4$ declines rapidly to $\chi\approx0$ near $h=\pm1$, which indicate the phase transitions. For comparison, we measure the Berry curvature $\mathcal{F}(\boldsymbol k)$ from the quasi-adiabatic dynamical response \cite{Schroer2014,Roushan2014,Tan2018,Tan2019} and obtain the Chern number $$C=\frac{1}{2\pi}\int_{T^2}\mathcal{F}(\boldsymbol k)dk_xdk_y\approx0$$ for all $h$ as excepted, which is plotted in Fig. \ref{TopTrans}(b).

\emph{Conclusion.}-- In summary, we have demonstrated two methods to measure the quantum metric using a superconducting qubit. We have further explored a topological phase transition characterized by the ECN in a simulated TRS system. The technologies developed here can be readily generalized to reveal the geometry and topology of other intriguing quantum states, such as tensor monopoles \cite{Palumbo2018} and non-Abelian Yang monopoles \cite{ColdAtom4} defined in four- and five-dimensional parameter spaces, respectively. Our work provides a novel way to experimentally explore the geometry and topology of the quantum states beyond the terminology of the Berry curvature.

\acknowledgments
The authors thank Q. Liu and G.Xue for technical support. This work was supported by the National Key Research and Development of China (Grant No. 2016YFA0301800), the National Natural Science Foundation of China (Grants No. 11604103, No. 11474153, No. 91636218,
and No. 11890704, No. U1830111 and No. U1801661), the Natural Science Foundation of Guangdong Province (Grant No. 2016A030313436), the Key Research and Development Program of Guangdong Province (Grant No. 2018B030326001), the Key Program of Science and Technology of Guangzhou (Grant No. 201804020055), and the Startup Foundation of South China Normal University.

X. T. and D.-W. Z contributed equally to this work.\\

\emph{Note added:} Recently, we became aware that two different experiments have been performed to measure the quantum metric tensor using an NV center spin in diamond \cite{MYu2018} and a two-dimensional photonic system \cite{Gianfrate2019}, respectively.


\begin{thebibliography}{99}
\bibitem{note1} By numerical simulations, we have checked that the sudden approximation is valid when the quench time $T\lesssim 10$ ns in the quench protocals. In our experiments, we have confirmed this condition and set $T=5$ ns.
\end{thebibliography}
\end{document}